\documentclass[conference]{IEEEtran}

\usepackage{graphicx}
\usepackage{lipsum}
\usepackage{multicol}
\usepackage{url}

\begin{document}
%
\title{Permissioned Blockchain-Based Security for SDN in IoT Cloud Networks}

\author{
	\IEEEauthorblockN{Safi Faizullah\IEEEauthorrefmark{1}, M.Asad Khan\IEEEauthorrefmark{2}, Ali Alzahrani\IEEEauthorrefmark{1}, Imdadullah Khan\IEEEauthorrefmark{3}
	}
	\IEEEauthorblockA{\IEEEauthorrefmark{1}\textit{Department of Computer Science,Islamic University,} \\ Madinah, Saudi Arabia
		\\ \IEEEauthorrefmark{1}safi@iu.edu.sa, alnaashi1@gmail.com}
	\IEEEauthorblockA{\IEEEauthorrefmark{2}
		\textit{Department of Telecommunication, Hazara University,} \\  Mansehra, Pakistan
		\\ \IEEEauthorrefmark{2}asadkhan@hu.edu.pk}
\IEEEauthorblockA{\IEEEauthorrefmark{3}
		\textit{Department of Computer Science, LUMS,} \\ Lahore, Pakistan
		\IEEEauthorrefmark{3}  imdad.khan@lums.edu.pk}
	
}


%


\maketitle

\begin{abstract}
The advancement in cloud networks has enabled connectivity of both traditional networked elements and new devices from all walks of life, thereby forming the Internet of Things (IoT). In an IoT setting, improving and scaling network components as well as reducing cost is essential to sustain exponential growth. In this domain, software-defined networking (SDN) is revolutionizing the network infrastructure with a new paradigm. SDN splits the control/routing logic from the data transfer/forwarding. This splitting causes many issues in SDN, such as vulnerabilities of DDoS attacks. Many solutions (including blockchain based) have been proposed to overcome these problems. In this work, we offer a blockchain-based solution that is provided in redundant SDN (load-balanced) to service millions of IoT devices. Blockchain is considered as tamper-proof and impossible to corrupt due to the replication of the ledger and consensus for verification and addition to the ledger. Therefore, it is a perfect fit for SDN in IoT Networks.  Blockchain technology provides everyone with a working proof of decentralized trust. The experimental results show gain and efficiency with respect to the accuracy, update process, and bandwidth utilization.
\end{abstract}


%
\IEEEpeerreviewmaketitle

\section{Introduction}
Networking technology has been continuously improving for the past many decades. This improvement has led to the emergence of cloud computing with massive computing and storage capacity in this era of Big Data. The widespread networking facilities and high-speed connectivity among devices form the backbone of the internet of things (IoT). According to Gartner, Inc.'s recent report, by $2020$ the enterprise and automotive Internet of Things (IoT) devices will expand to $5.8$ billion, which is a $21\%$ increase from 2019 \cite {r1}. The cloud networks managing IoT infrastructure will grow in size and sophistication to cater to the growing computational demands. To overcome the issue of diverse geographical spread, distributed IoT network offers promising solutions. However, still at the core legacy networks, composed of hardware-based, which are costly, inflexible, and difficult to scale and manage. An approach Software-defined networking (SDN) is proposed to cater to different challenges caused by complex and large network and enables comprehensive network programmability and improve load management. 
\begin{figure}[h]
	\centering
\includegraphics[scale = 0.6, page = 1]{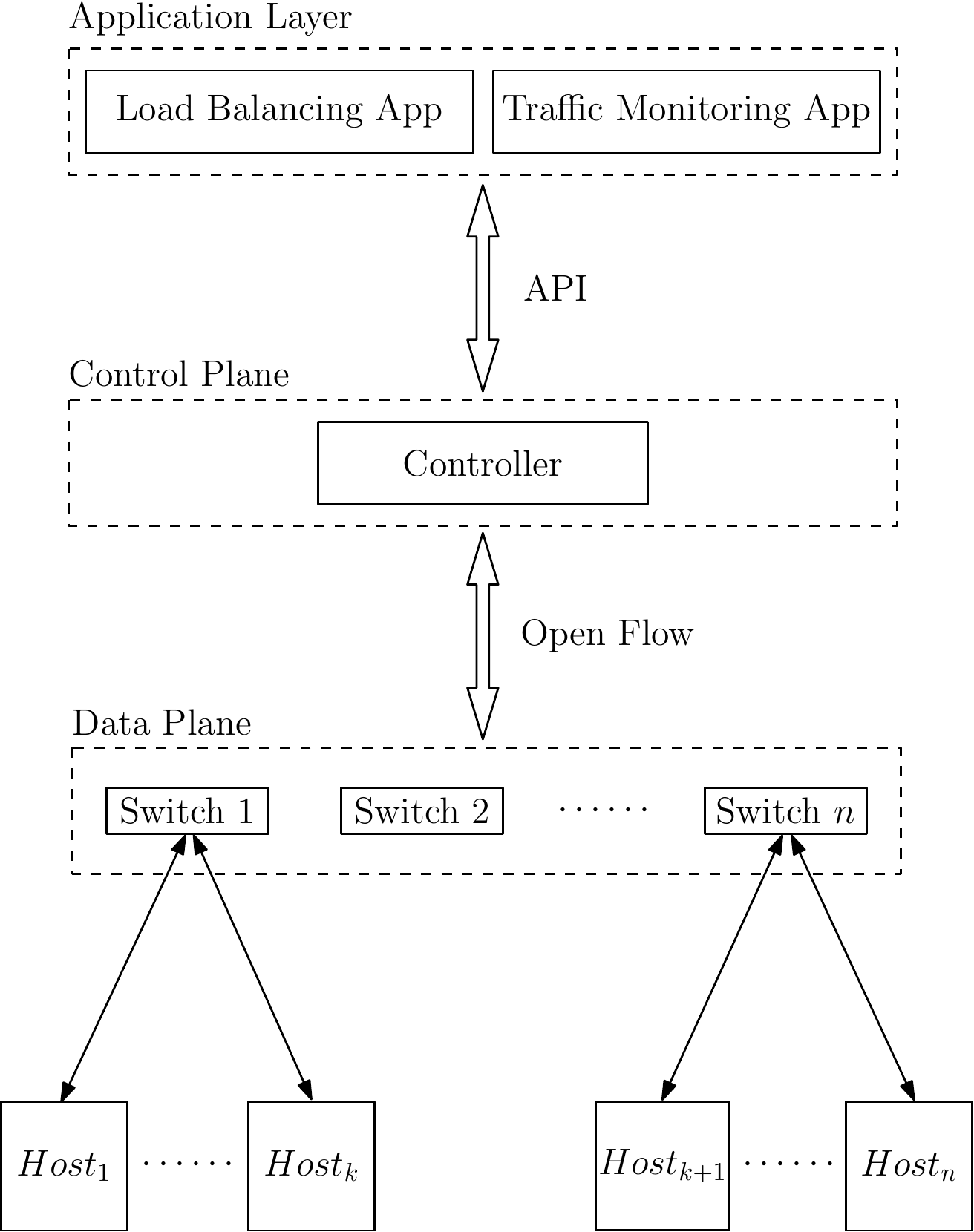}
\caption{SDN Architecture}
\label{fig_sdn_architecture}
\end{figure}

In SDN architecture, the data and control planes are divided into two separate layers, which simplifies network management. The control plane consists of a controller whose responsibility includes networking intelligence and policy-making. Although the division of layers allows the scalability of the network, however, it also introduces new vulnerabilities. In a system consisting of multiple SDN networks, problems like security, performance, reliability, and scalability arise due to the centralized control architecture. We have identified certain scenarios in which the SDN architecture becomes vulnerable to the attackers. These vulnerabilities allow attackers to execute a distributed denial of service (DDoS) attack on the network \cite{r2}. The DDoS attack can be performed by repeatedly sending unique packets requests to the controller. Many existing studies highlight security and other issues with SDN. To address these issues, a distributed architecture (which contains multiple SDN controllers connected to each other) is introduced \cite{r3,r4}. Several existing studies focus on the problem of state consistency among numerous controllers. In the distributed architecture, the flow of data between the controllers and the forwarding devices is configured statically. This static configuration can result in an uneven distribution of loads between the controllers. In addition to these problems, a distributed SDN network is required, which possess properties like high availability and low response time. Although there are multiple studies in the literature offering a reliable and scalable solution to the distributed network for management \cite{r1, r3, r5}, however, none of these studies have solved this problem completely.

Blockchain (a distributed data storage approach) has recently drawn attention from researchers, which is used to solve the existing problems of SDN architecture \cite{r6, r7}. Using the blockchain approach, we can manage the applications in a distributed manner, which was run previously through a trusted intermediary. All functionalities of the central approach can be achieved without the need of any central authority. The blockchain approach offers a distributed peer-to-peer network in which the untrusted individuals can interact with each other in a verifiable manner without the need of trusted intermediary \cite{r8,r9}. Permissioned blockchain utilizes access control to alleviate the burden of proof-of-work by assigning certain roles to known entities \cite{r10}.

In this paper, we offer a blockchain-based solution that is provided in redundant SDN as a load-balancer to service millions of IoT devices, on an execute-order architecture that supports high throughput of transactions. The IoT devices will communicate and get connected to a blockchain-based SDN network that it utilizes. Our main contributions are following:
\begin{enumerate}
	\item We define an execute-order architecture for blockchain-based security for SDN in cloud networks based IoT setting.
	\item We extend permissioned blockchain architecture for IoT device communications and give a proof-of-concept implementation of the proposed system.
\end{enumerate}

\section{Background and Related Work}
In this section, we provide a summary of the background and related work related to SDN, Blockchain, and Access Control in Permissioned Blockchain that we use in our proposed work. We represent the basic Software-Defined Networking (SDN) architecture in Figure \ref{fig_sdn_architecture}. From top-down, we have applications that use the two layers of control plane and data plane for communications. Note that the switches in SDN do not construct the forwarding table automatically but instead rely on the controller to construct a flow table \cite{r2,r3}. The control plane of the SDN architecture consists of a centralized controller, which is the most important new feature provided by SDN. 

The centralized controller help in network management, and configuration information is stored in a single place instead of distributing it to individual network devices. Additionally, the controller has a centralized view of the network, which makes it capable of calculating optimal routes across the network, constructing flow tables, and insert them into each of the switches without relying on distributed routing algorithms. Lastly, the application layer in SDN allows applications to request specific network behavior from the controller. As SDN is taking share and several protocols are being studies, OpenFlow is emerging as one of the most used protocols in SDN. OpenFlow \cite{r6} is defined as an open standard and as such not tied to any single controller, making it suitable for heterogeneous networks. It is a preferred protocol that is being utilized between the controller and SDN switches. It transmits messages from the controller to the switch to facilitate the flow tables constructions/updates. Also, it transmits messages from the SDN switch back to the controller so that switches can inform the controller about the various events in the network, such as new host or switch additions and other similar functions. OpenFlow protocol supports three types of messages: controller-to-switch, asynchronous, and symmetric messages. The controller initiates Controller-to-switch messages, while asynchronous messages are messages sent by the switch without being requested by the controller. These messages include packets arrival, switch state changes, error states, etc. Lastly, symmetric messages are the messages sent without solicitation, in either direction. In this new architecture, every SDN switch consists of a flow table, which in turn stores a set of flow entries. Whenever the switch processes a new packet, it is compared against the flow table to find a matching entry. Once a matching entry is located, actions specified for that entry are performed on the packet e.g. forward a packet out a specified port. If no matching entry is found, the packet is forwarded to the controller, which can decide the action to be taken on the packet. Each flow table entry consists of a rule or header field to match against the incoming packet. Note that every flow entry is associated with zero or more actions that dictate how the switch handles matching packets. If no actions are specified for an entry, then such a packet must be dropped. Flow entry also contains counters that are maintained per-table, per-flow, per-port, and per-queue.

In recent years, blockchain is being used in many domains besides the financial setting. It is a perfect fit for use in SDNs. Several studies in the literature have designed the Access control mechanisms using the blockchain approach \cite{AafafPrivacy2017, AafafFairAccess2016, XuBlendCAC2018}. However, these approaches fail in terms of security, scalability, and efficiency in most cases.
\begin{figure}[ht]
\begin{center}
\includegraphics[scale=0.6, page = 1]{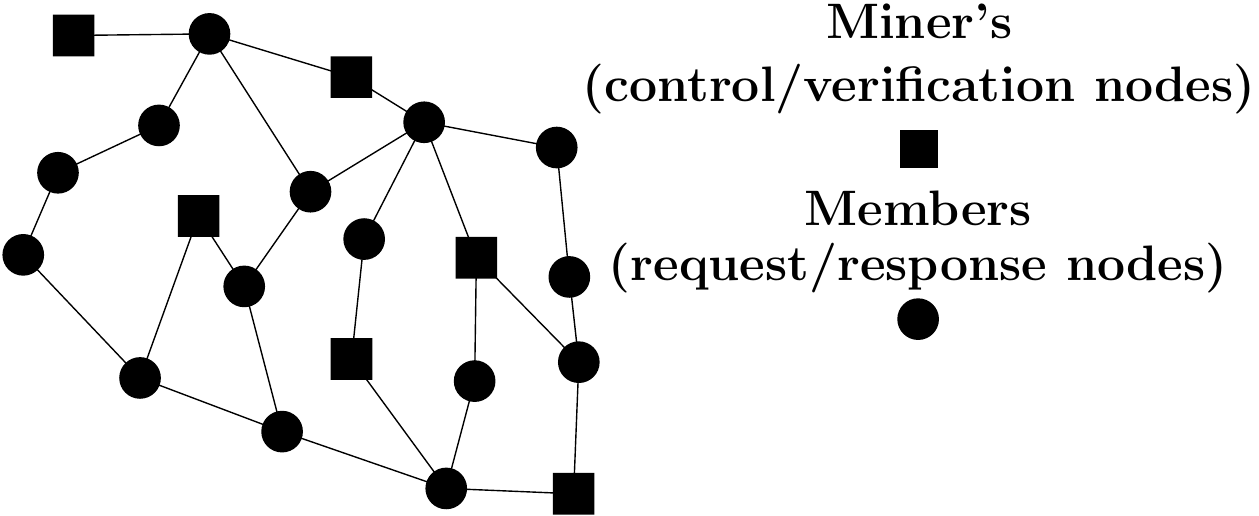}
\end{center}
\caption{Blockchain Network of $21$ nodes with $15$ Request/Response and $6$ Controller/Verifier nodes}
\label{fig_blockchain_network}
\end{figure}

\section{Blockchain Architecture}
A blockchain is a secure public ledger to store transactions. The blockchain does not have a central storage system instead all the transactions are replicated at the user end (nodes). Each user has the same copy of the ledger, and each user keeps on updating its copy hence ensuring complete consensus from all users. At the same time, the blockchain technology provides everyone with a working proof of decentralized trust. To make a transaction, it is necessary that the user has an updated copy of the ledger. The transactions are grouped into blocks, and after a specific timestamp, a block (consisting of multiple transactions) is published. Publishing a new block means that the new block is appended at the end of the chain of blocks. In addition to this, it is mandatory that before appending a new block in the chain, all transactions in the last block of the chain must be verified. The verification of transactions means to check that all transactions are legitimate. This implies that appending a new block in the chain also proves that the transactions done earlier are all legitimate. By doing so, blockchain is termed to be tamper-proof and impossible to corrupt. See Figure \ref{fig_blockchain_network} for a reference blockchain architecture.

\section{Permissioned Blockchain }
Access control mechanisms are widely using blockchain based solutions to manage access to sensitive and personal data \cite{r2,r11}. Access control systems manage the access of sensitive and personal data. Attribute based Access Control (ABAC) define an access control paradigm by which access rights are given to the requesting entity (i.e. user) using the policies which combine attributes together.
Transferring access rights means to sell or grant access
rights to other users. In case of an access request, the requester should have a mechanism to verify the decision of access request. 
In this paper, we propose an approach based on blockchain, which allow the transfer of access rights among requesting entities and also represent the right to access a resource. A blockchain is a distributed and temper resistant database which allows append only operation. In a Blockchain, blocks are maintained by saving the link to the previous block where each block contains a set of non-conflicting transactions. In this paper, the blockchain technology is used to represent the rights to access resources and to transfer rights from one user to another. The resource owner specifies the access rights (policies) on its resources and each policy is stored as a transaction in the blockchain called Policy Creation Transaction (PCT). In short, resource owner, through policies, specifies the user who can have access to its resources and what set of transactions is required for the access grant. The owner can modify the conditions specified in a policy anytime. Using blockchain as proposed, the right to access a resource can easily be transferred from the owner to another user. A Right Transfer Transaction (RTT) tracks the policy whose right are exchanged. In RTT the parties involved are the current right holder and the object to which the rights are being transferred to and the intervention of the owner is not required \cite {r12}.


The proposed model is deployed on the Bitcoin blockchain by developing a proof of concept implementation scheme. Operations associated with policies like creation, pupation etc. are performed using the Bitcoin transaction. Two commonly used methods based on the Bitcoin scripting language to store arbitrary data on the blockchain. To embed data in a transaction, first a Bitcoin transaction is created using tokens which is the value used in transactions.
A policy is created by a resource owner through a new Bitcoin transaction having inputs and outputs. The outputs create tokens to do rights exchange and to update the policy. The exchange of rights between two subjects takes place when subjects are interested in buying or selling the rights. The right exchange is done by a message change protocol to jointly build and sign the RTT. As each policy and its updates are publicly visible in blockchain, the right associated with a policy can be verified by any subject. The approach mentioned in the paper is validated through a reference implementation based on the Bitcoin. Access control has been one of the major challenges in the Internet of Things domain \cite{r13}. With the evolution of IoT
field, an effective mechanism for access control (authentication and authorization methods) to the private and sensitive data is of vital importance. This survey paper compares access control environments in IoT and describes models, prototypes and protocols in IoT. The paper describes different access solution in Objectives, models, Architecture and Mechanism (OM-AM) way. An access solution has three components
authentication, authorization and accountability. The paper focuses on authentication which is to define a security policy and to implement a model encapsulating that policy. 

\section{SDN Flow Rules Tables in Blockchain Network}
The objective in this context is to make an access control policy that define rules related to the regulation of access control. The authorization level of OM-AM model bridges the gap between high level policies to low level mechanisms of how authorization rules will be applied. Architecture describes the participating entities, the work flow and interactions. Mechanisms define the low-level functions to enforce policies and how access requests are evaluated against those policies. In an access control solution, three main constraints that need to be dealt are flexibility, scalability and heterogeneity. In addition to this, access control solutions are: a) role-based b) attribute-based c) capability based and d) organizational based. The analysis and evaluation of the existing works are also highlighted. The paper then discusses some open challenges in the authorization and access control which are i) existing access control mechanisms vs new mechanism keeping in view IoT specific requirements ii) centralized vs distributed access control policies. In the end paper presents some research trends and gives future directions in IoT. Access control is a big challenge in IoT devices as it is computationally infeasible to implement access control mechanism on a device with limited computational capacity. Managing access control through a third party may harm user privacy as the private information can be breached through some security flaw. The paper devises a blockchain based decentralized privacy preserving authorization framework, FairAccess, to control access and make authorization decisions. The framework operates by defining a digital signature called Authorization Token which controls the access to specific resource. FairAccess stores the access control polices between the resource and the requester as transactions in the blockchain. It provides required authorization functionalities like grant access, revoke access, request access etc. The paper also discusses the validation of access control policy and the detection of token reuse.

\section{Blockchain-Based Proposed Approach}
In this section we are presenting a Blockchain based security and access control mechanism for IoT network.

\subsection{IoT Network model }
We are considering an IoT network based on the thin client model. In this model IoT devices will be without any processing capabilities as known as Thin client and will depend on its server for any processing. As this setup is following centralized network model that will provide greater flexibility and programmability. But on the other side this centralization also come with problems like scalability and single point of failure etc. Therefore, a physically distributed architecture with virtually centralized control is required for optimal results. As there will be different servers, each managing a group of IoT thin client devices. Further each server will be access by different users according to the IoT device it controlling and the type of task (operation) user required. Now the problem is that all these servers should maintain same state (i.e. same policy and status of network) without any centralized entity. Blockchain is the most suitable solution for this purpose. As it facilitates security and access control for whole network without centralized entity. In out setting, all the servers will be part of blockchain where each server will maintain a database (copy of open ledger). This database will store information about the access policy and security mechanism. Further, each change will be considered as a blockchain transaction. To add a transaction in blockchain, it should be validated by servers.

Let consider $ R_{i j} ^k$ is a request from $ith$ users direct to the $jth$ IoT device and carrying the $kth$ task. This request must be from a valid user, must be directed to an IoT device that access is allowed to this user and should carry a type of task that is permissible for the user and device both. Additionally, the network (blockchain) must ensure that no device should operating within its maximum capacity (i.e. ensuring load balancing and avoiding target security attacks).

In the following subsections we are presenting security, load balancing and access control mechanism in detail.

\section{SDN Flow Rules Tables in Blockchain Network}
\subsection{Security}
There are different types of security attacks that can be carried on an IoT network. 
 Most importantly, user validity should be check i.e. if the user is allowed to access any device in the network.
Using the blockchain technology, access policy details will be available at each server. In case of a request from a user, the server will check its data base for user validation and also share this request in the blockchain.
Each server will compare this request with the access policy. In case the user detail (different techniques are used for defining access policy e.g. Attribute Based Access Control) are matching with the access policy. The servers in the blockchain will verify this request, send verification to the target (local) server and add this transaction in the blockchain.

\subsection{Load Balancing}
This is mechanism to avoid overloading a IoT device by controlling/migration number of users (requests) per device in an unit time.
 Let a user send a request for a specific task on an IoT device, the permission for this operation should be
conditioned to the current load on that device.
To ensure this each request will be forwarded to the blockchain by the local server (also checking by itself). Each server will check the current load on the target device and will calculate the total load on the device if the current request is allowed.
 Each server will send back its decision about the current request to the local server. In case the request is allowed/verified, it will be added to the blockchain and also update the device load.
Further, this mechanism avoids and protect the network from overloading even if the local server act in greedy manner or compromised in result of an attack.
\begin{figure}[t]
\centering
\includegraphics[scale = 0.4, page = 6]{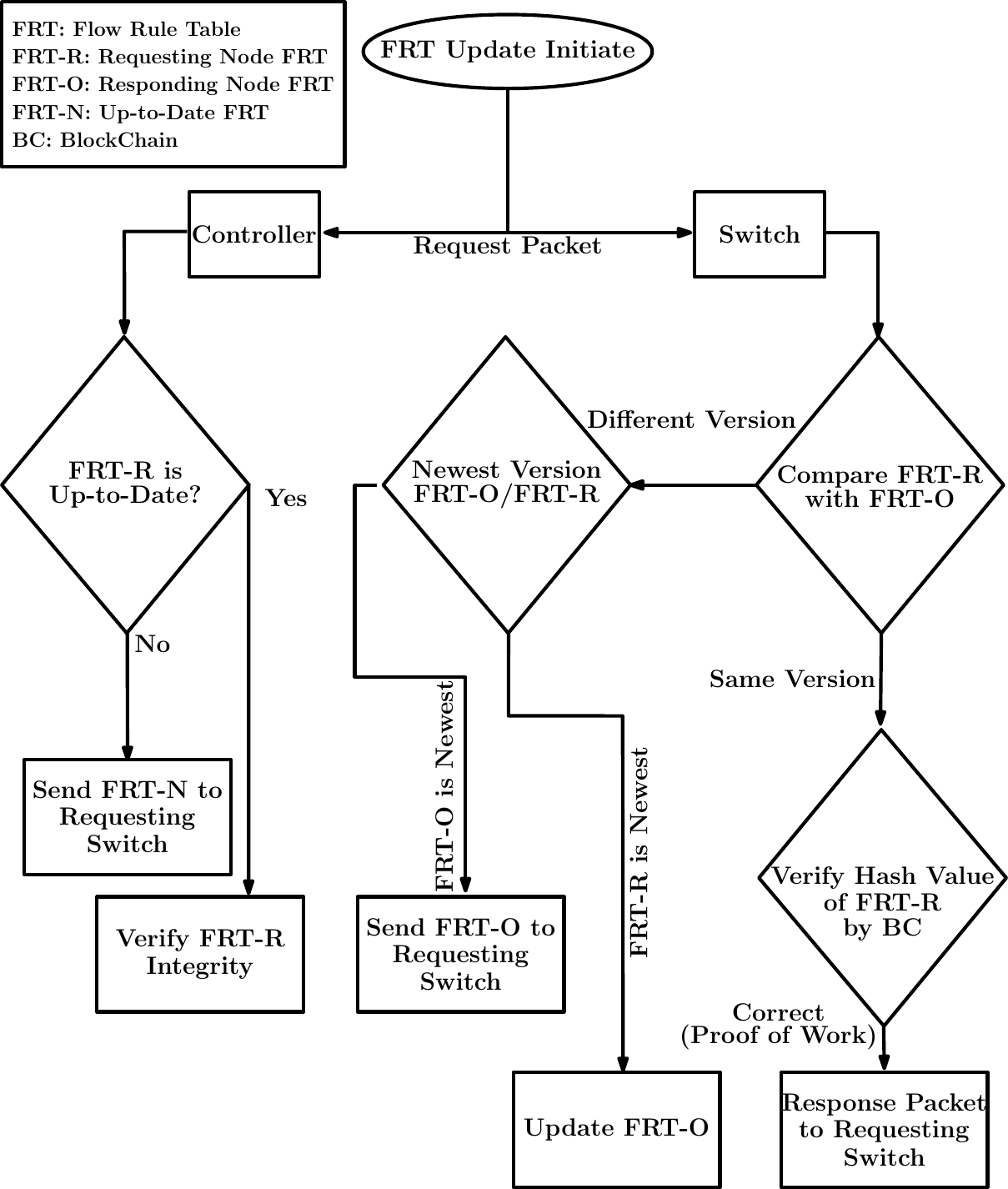}
  \caption{Flow Diagram of Flow Rules Table Update Process}
  \label{fig_flow_chart}
\end{figure}

\subsection{Access Policy Implementation}
Considering an industrial IoT network (IIoT), where different type of device work together. Each device can perform different level of operations. Similarly, users in IIoT can be divided in to different levels and each type of users are authorized to access/request/change specific task/policy.
To ensure the above access granularity for each device, each task will be first validated from the blockchain (even from the authorized users request).
This verification through blockchain will ensure that even a valid user is blocked to perform a task (operation) that is not allowed to him/her.

Figure \ref{fig_blockchain_network} depicted an overview of an IoT setup that is built by utilizing the concept of  blockchain SDN, consists of miners  (controller / verification) nodes serving the connected members (switches / IoT forwarding devices). In the rest of this work, we will use the controllers and switches as general terminology. Controllers are mainly responsible for functions related to flow rule tables such as storing, updating these tables and processing switches queries. The main function of switches is to forward the user's data according to rules defined in the flow tables. Further, both the members (switches) can operate in requesting or responding modes. When a switch sends a request e.g. a table update process initiation, this switch is in the requesting mode while all the other nodes (both switches and controllers) will be considered in the responding mode.


The IoT switch enters in request mode when it broadcast a packet request with a version check to update its flow rules table. The response to the switch depends on whether the responding node is a switch or a controller, a detailed process of response is presented in Figure \ref{fig_flow_chart}. In the following, we will explain the response from both switches and controllers. First, we will consider the case if the receiver of the request packet is a controller:

\begin{itemize}
\item {The controller verifies that the requesting flow rule table (FRT-R) is an update-to-date version or not.}
\item { In case the FRT-R is not an up-to-date copy, the controller will send the latest copy of FRT to the requesting switch.}
\item {Otherwise, controller will check the integrity of FRT-R.}
\end{itemize}




In case the receiver is a switch then

\begin{itemize}
\item {First, it checks the version of the requesting switch FRT i.e. FRT-R by comparing it with its own flow table i.e. FRT-O}
\item {If both FRT's have the same version then the responding switch sends a request to other nodes in the blockchain to calculate the hash value of the FRT-R.}
\item {If the responding switch received a correct hash value (proof of work) from the blockchain, responding switch will confirm FRT-R via a response packet.}
\item {In case the FRT-R and FRT-O have a different version of FRT then the latest FRT will be determined between FRT-R and FRT-O.}
\item {If the FRT-R is the latest version compare to FRT-O, responding switch will update FRT-O.}
\item {Otherwise, the responding switch will send FRT-O to requesting switch via response packet.}
\end{itemize}


\section{Performance Evolution}
In this section, we present the implementation detail and experimental results of our proposed architecture. We performed extensive experiments to evaluate our approach in terms of accuracy, scalability, security, and efficiency.


\begin{figure}[ht]
\begin{center}
\includegraphics[scale = 0.85, page = 4]{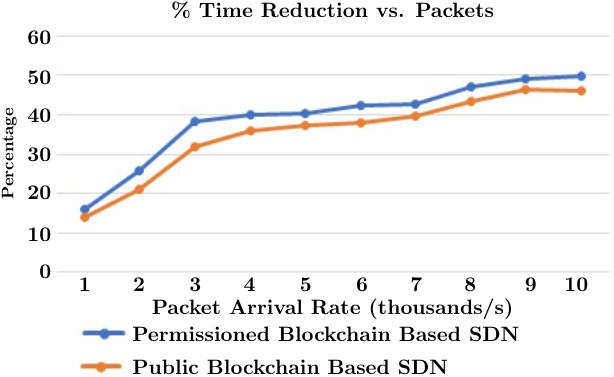}
\end{center}
\caption{Flow Update Time Performance as Percentage for Permissioned Blockchain Based SDN vs. Public Blockchain Based SDN}
\label{Percent_time_reduction}
\end{figure}

\begin{figure}[ht]
\begin{center}
\includegraphics[scale = 0.85, page = 5]{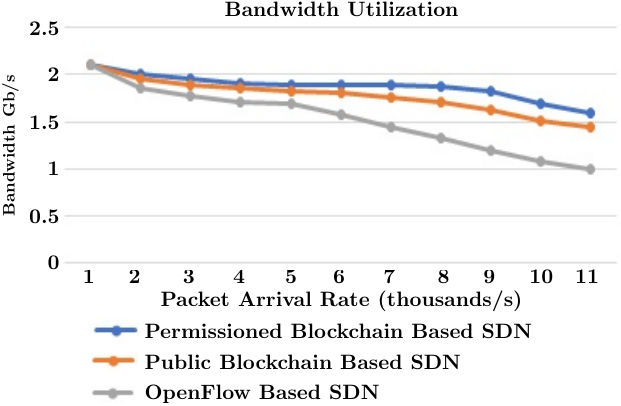}
\end{center}
\caption{Bandwidth Utilization comparison for different SDN Architectures}
\label{bandwidth_utilization2}
\end{figure}

A CPU based cluster of 15 Intel i7 3.40 GHz with 32 GB RAM servers and SDN based blockchain network with 10 controllers/verifications and 990 request/response nodes. Figure \ref{fig_blockchain_network} represents a sample of this experimental setup. In order to achieve scalability we choose OpenFlow software switch and OpenVSwitch. Next, we built a normal distributed SDN to compare the flow rule table update performance of our proposed Blockchain IoT architecture in large-scale network. 
Flow rule tables update time is critical for the smooth and efficient operation of IoT network. Figure \ref{Percent_time_reduction} depicts the comparison results of permissioned Blockchain based SDN and public Blockchain based SDN. It can be seen from the experimental results that the performance of our proposed architecture (permissioned Blockchain based SDN) continuously improves as compare to the public blockchain based SDN in terms of percentage time reduction as the rate of packet-in arrival increases. 
Moreover, Figure \ref{bandwidth_utilization2} shows the bandwidth utilization comparison of different SDN architectures.  
Figure \ref{bandwidth_utilization2} provides the comparative bandwidth utilization of permissioned, public and OpenFlow SDNs.

The defense effectiveness of our approach is measured using bandwidth utilization. We compare the bandwidth utilization of our approach with the OpenFlow network (not utilizing blockchain).
For the software test environment, the MININET SDN emulation tool is used. The clients were deployed to send a DoS attack (UDP floating) to the switches, and bandwidth utilization of clients is measured with and without the DoS attack. Figure \ref{bandwidth_utilization2} provides the comparative bandwidth utilization of permissioned, public and OpenFlow SDNs.  In the start (when there is no attack), the bandwidth starts is $2.1$ Gb/s. As the number of attacks increases, the bandwidth decreased rapidly. The bandwidth is almost half when the packet arrival rate reaches $1000$ packets per second. As we triple the number of packets/s, the whole network started coming down and malfunction. When using the proposed architecture, the bandwidth starts at $2.1$ Gb/s and stays stable even when traffic is tripled from $1000$ packets/s. 

\section{Conclusion}
In this paper, we propose an innovative, secure SDN in IoT network architecture using permissioned blockchains. We analyze the challenges faced by large scale IoT networks after the introduction of new communication paradigms. Our approach addresses those challenges effectively and performs well under the scenarios where the previous methods failed to show efficient results. The system improves the network’s performance and capacity in the face of network attacks such as spoofing and DDoS/DoS attacks. Our approach reduces the response time of the attack by allowing IoT forwarding devices to check for any change in the latest flow rules table. We evaluate our approach (permissioned blockchain based SDN) in terms of performance overhead, scalability, and bandwidth utilization efficiency and compare it with Public blockchain based SDN and  OpenFlow SDN models. The experimental results show that our proposed method outperforms the existing baseline models. In the future, we will extend our proposed model to work with distributed cloud computing architectures in order to evaluate and compare it with other proposed solutions.


\section*{Acknowledgment}
This research is partially supported by the Research Deanship of Islamic University of Madinah. 

\bibliographystyle{IEEEtran}
\bibliography{Permissioned_Blockchain_Based_Security}

\end{document}